\documentclass[a4paper, 11pt]{article}
\usepackage{epsf}
\usepackage{epsfig}
\usepackage{fancyhdr}
\usepackage{amsmath}
\usepackage{amssymb}
\usepackage{theorem}
\numberwithin{equation}{section}

\begin{document}
\title{What can we learn from noncoding regions of similarity between genomes?}
\author{Thomas A. Down and Tim J. P. Hubbard \\
{\it \{td2,th\}@sanger.ac.uk}}
\date{\today}
\maketitle

\begin{abstract}
{\bf Background:} In addition to known protein-coding genes, large amount of apparently non-coding
sequence are conserved between the human and mouse genomes.  It seems reasonable
to assume that these conserved regions are more likely to contain functional
elements than less-conserved portions of the genome.  Here we used a motif-oriented
machine learning method to extract the strongest signal from a set of non-coding
conserved sequences.

{\bf Results:}We successfully fitted models to reflect the non-coding sequences,
and showed that the results were quite consistent for repeated training runs.
Using the learned model to scan genomic sequence, we found that it often
made predictions close to the start of annotated genes.  We compared this method 
with other published promoter-prediction systems, and show that the set of promoters
which are detected by this method seems to be substantially similar to that 
detected by existing methods.

{\bf Conclusions:} The results presented here indicate that the promoter signal
is the strongest single motif-based signal in the non-coding functional fraction
of the genome.  They also lend support to the belief that there exists a substantial
subset of promoter regions which share common features and are detectable by
a variety of computational methods.
\end{abstract}

\section{Background}

Since the publication of draft sequences for the human \cite{human.genome} and 
mouse \cite{mouse.genome} genomes, several groups have run large-scale comparisons
of the sequences to detect regions of conserved sequence.  An initial
survey of these was published along with the draft mouse genome \cite{mouse.genome}. 
Briefly, protein coding genes are -- as we might expect -- among the most strongly conserved
regions, but homologous sequences can be found throughout the genome. In total, it is
possible to align up to 40\% of the mouse genome to human sequence \cite{schwartz.blastz}, but it seems
likely that at least some of this is just random ``comparative noise'' -- regions of
sequence which serve no particular purpose but which, purely by chance, have not
yet accumulated enough mutations to make their evolutionary relationship unrecognizable.  However, it 
is widely accepted that
some of the noncoding-but-similar regions, especially those with the highest
levels of sequence identity between the two species, are preferentially conserved
because they perform some important function.  It has been estimated that around
5\% of the genome is under purifying selection \cite{mouse.genome}, indicating that
mutations in these regions have deleterious effects: a strong suggestion of some
important function..

Here, we apply the Eponine Windowed Sequence (EWS) sequence analysis method \cite{down.rvmseq}
method which uses a Relevance Vector Machine \cite{tipping.rvm} to extract a
minimal set of short motifs which are able to discriminate
between two sets of sequences: in this case, a positive set of conserved non-coding
sequences and a negative set of randomly picked sequences.  The EWS model is
an adaption of the Eponine Anchored Sequence model first described in \cite{down1}
and subsequently used to predict a range of additional biological features including
translation start sites and transcription termination sites [A. Ramadass, unpublished]
While EAS is designed to classify individual points in a sequence -- a feature
which allows the EponineTSS model to predict precise locations for transcription
start sites -- EWS classifies
complete blocks (windows) of sequence.  The design and implementation of the EWS
model is described in detail in \cite{down.rvmseq}.

\section{Results}

\begin{figure}[!bth]
\begin{center}
\includegraphics[scale=0.66]{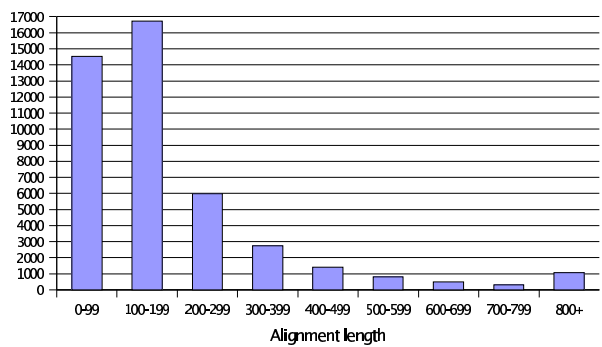}
\caption{Number of blastz alignments of specific lengths to human chromosome 6}
\label{length.histogram}
\end{center}
\end{figure}

We considered a set of `tight' alignments made by the blastz program \cite{schwartz.blastz} between
release NCBI33 of the human genome and release NCBIM30 of the mouse genome.  In total,
this method reported 787173 blocks alignable between the two genomes.  We considered
only those blocks assigned to human chromosome 6, a 170Mb chromosome which
has recently undergone manual annotation of gene structures and other features 
\cite{mungall.chr6}.  This chromosome included 44105
(5.6\%) of the total alignments.  These varied in length from 34 to 9382 bases,
with a length distribution skewed towards relatively short alignments, as shown in figure
\ref{length.histogram}.

Since we were interested in non-coding features of the genome, we ignored all
regions where an alignment overlaps an annotated gene structure.  This
removed 20.8\% of aligned bases.  It is
possible that some genes, and especially psuedogenes, have been missed by the
annotation process, so we also removed portions covered by {\it ab initio} gene
predictions from the Genscan program \cite{burge.genscan}.  This eliminated
an additional 4.3\% of aligned bases.
Finally, repetitive sequence elements annotated by the programs RepeatMaster \cite{smit.repeatmasker} and
trf \cite{benson.trf} (5.9\%) were removed from the working set.  The remainder of the aligned
regions were split into non-overlapping 200 base windows, ignoring any portions
less that 200 bases.  This gave a set of 13925 sequences which are well-conserved
between human and mouse -- and therefore likely to be functional --
but which are very unlikely to be part of the protein-coding repertoire.  These
formed the positive training set for our machine learning strategy.

A negative training set of equal size was prepared by picking 200-base windows
at random from the non-coding, non-repetitive portions of chromosome 6, using
the same criteria to define repeats and coding sequence.  While
it is probable that this set also included some functional sequences, we would
expect them to be represented at a substantially lower level than in the
conserved set.

These two sets of sequence were presented to the Eponine Windowed Sequence
machine learning system.  Randomly chosen 5-base words were used as seed motifs,
and three models were trained, each for 2000 cycles.  The set of motifs used
in model 1 is shown in table \ref{model.motifs}

\begin{table}[!bth]
\begin{center}
\begin{tabular}{p{3cm} | p{3cm} || p{3cm} | p{3cm}}
\multicolumn{2}{c||}{Positive} & \multicolumn{2}{c}{Negative}\\
Forward & Reverse & Forward & Reverse\\
\hline
gtca	&	tgac	&	tacgt	&	acgta\\
tattg	&	caata	&	gggca	&	tgccc\\
tgcca	&	tggca	&	gtca	&	tgac\\
ggca	&	tgcc	&	acaat	&	attgt\\
tacgt	&	acgta	&	ggggc	&	gcccc\\
gtact	&	agtac	&	tact	&	agta\\
taac	&	gtta	&	cctcc	&	ggagg\\
ttt	&	aaa	&	ggca	&	tgcc\\
acaat	&	attgt	&	tattg	&	caata\\
caatt	&	aattg	&	tattg	&	caata\\
cagc	&	gctg	&	aaatt	&	aattt\\
cag	&	ctg	&	caat	&	attg\\
cggat	&	atccg	&	gtat	&	atac\\
aaatt	&	aattt	&	ccagg	&	cctgg\\
gctcg	&	cgagc	&	catg	&	catg\\
ggc	&	gcc	&	act	&	agt\\
	&		&	taagg	&	cctta\\
	&		&	aaaaa	&	ttttt\\
\end{tabular}

\caption{Motifs used in EWS homology model 1.  The entries in this table show
  consensus sequences of the weight matrices used in the model (note that it
  is possible for two distinct weight matrices to have the same consensus
  sequence).  Motifs are listed in both forwards and reverse-complement orientation,
  and the two sections of the table indicate whether that motif is given a
  positive or negative weight in the learned linear model.}
\label{model.motifs}
\end{center}
\end{table}

\begin{figure}[!bth]
\begin{center}
\includegraphics[scale=1.0]{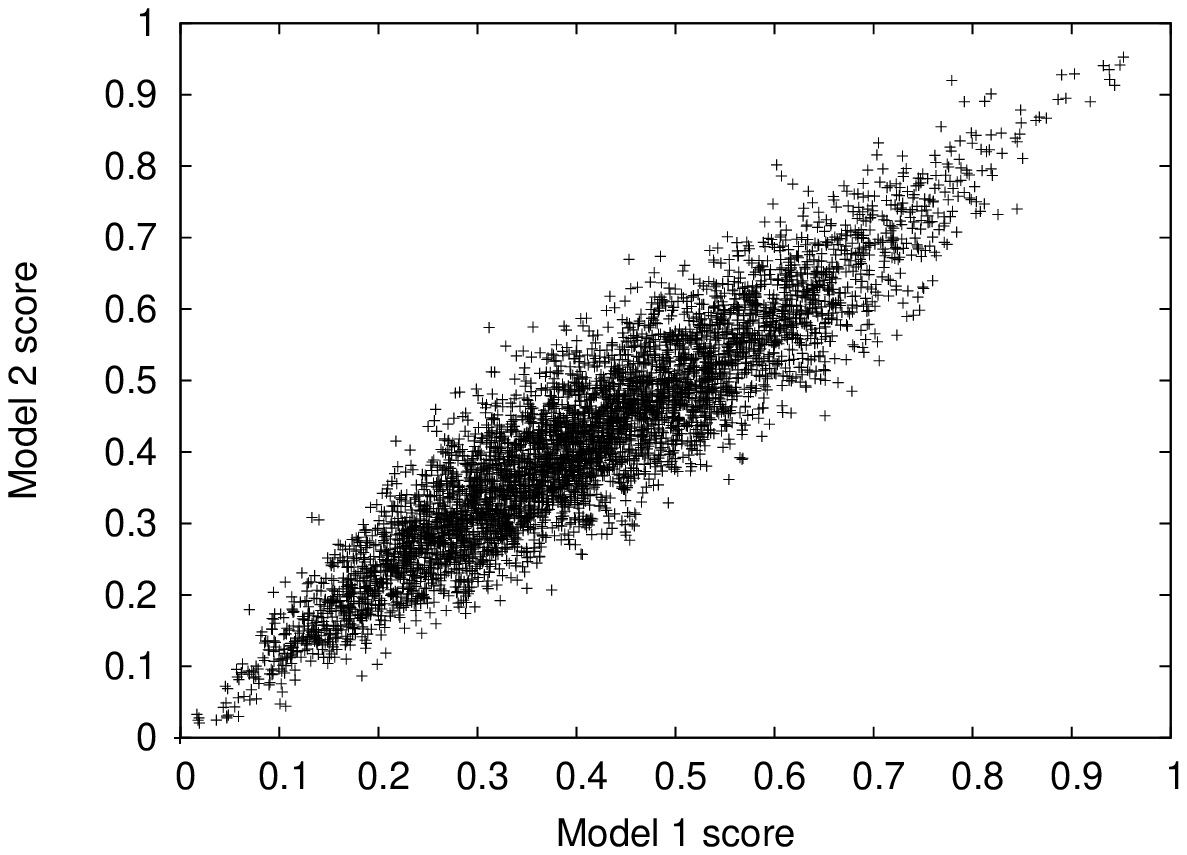}
\caption{Scatter plot showing the scores of models 1 and 2 on a set of sequences}
\label{scan.scatter}
\end{center}
\end{figure}

While the exact set of motifs used in the model varied
somewhat from run to run, testing pairs of models on non-overlapping windows
from a 1Mb region of human chromosome 22 and plotting the scores showed that
the model outputs were highly correlated ({\it e.g.} figure \ref{scan.scatter}). 
We calculated the Pearson correlation coefficient for all pairs, and in all cases
this was greater than $0.96$.  From this strong correlation, we concluded that
any variations in the model were simply the result of the trainer picking one
representative from a group of motifs which provide similar information.

\begin{figure}[!bth]
\begin{center}
\includegraphics[scale=1.0]{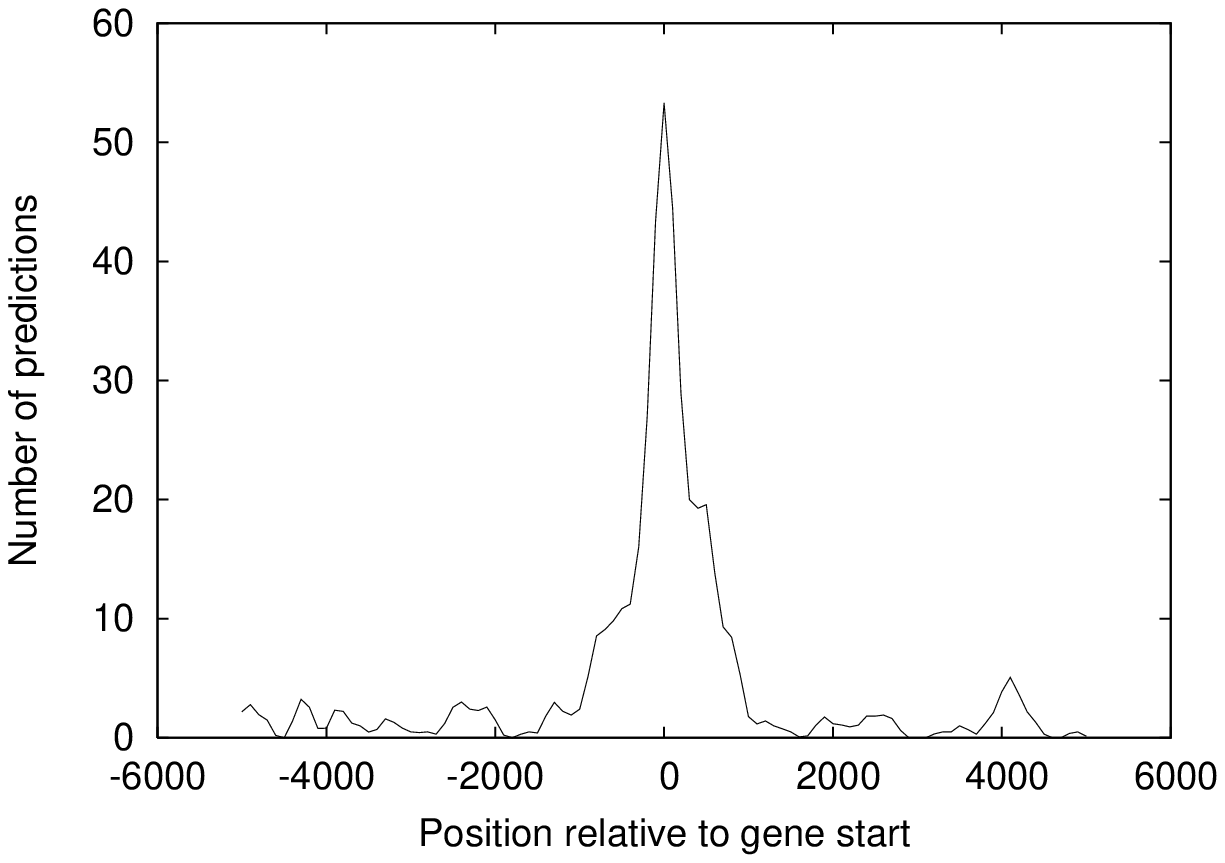}
\caption{Density of predictions from one of the homology models around known gene
  starts on human chromosome 22}
\label{c22.density}
\end{center}
\end{figure}

We scanned genomic sequences using these models at a range of thresholds,
and examined the results using the Ensembl genome browser \cite{hubbard.ensembl}.
Visual inspection showed that
many of the highest-scoring regions were  localized near the start of genes.
This prompted us to look at the distribution of high-scoring with respect to
the starts of a set of well-annotated genes.  We considered the GD\_mRNA genes
from version 2.3 of the human chromosome 22 annotation.  These are confidently
annotated genes with experimental evidence as described in \cite{collins.c22},
which confirms at least the approximate location of the ends of the transcripts, and
are entirely independent from the chromosome 6 training data.
Figure \ref{c22.density} shows the density of predictions with GLM scores $\geq0.90$ relative to the
annotated 5' ends of these genes.  This shows a strong peak of predictions close
to the annotated starts, demonstrating that the model is predicting some sequences
commonly located around the transcription start site of genes.  Combining this
observation with the fact that the model was trained from conserved (and therefore
presumed functional) sequences, we believe that it is detecting signals found
in the promoter regions of genes.

\begin{figure}[!bth]
\begin{center}
\includegraphics[scale=1.0]{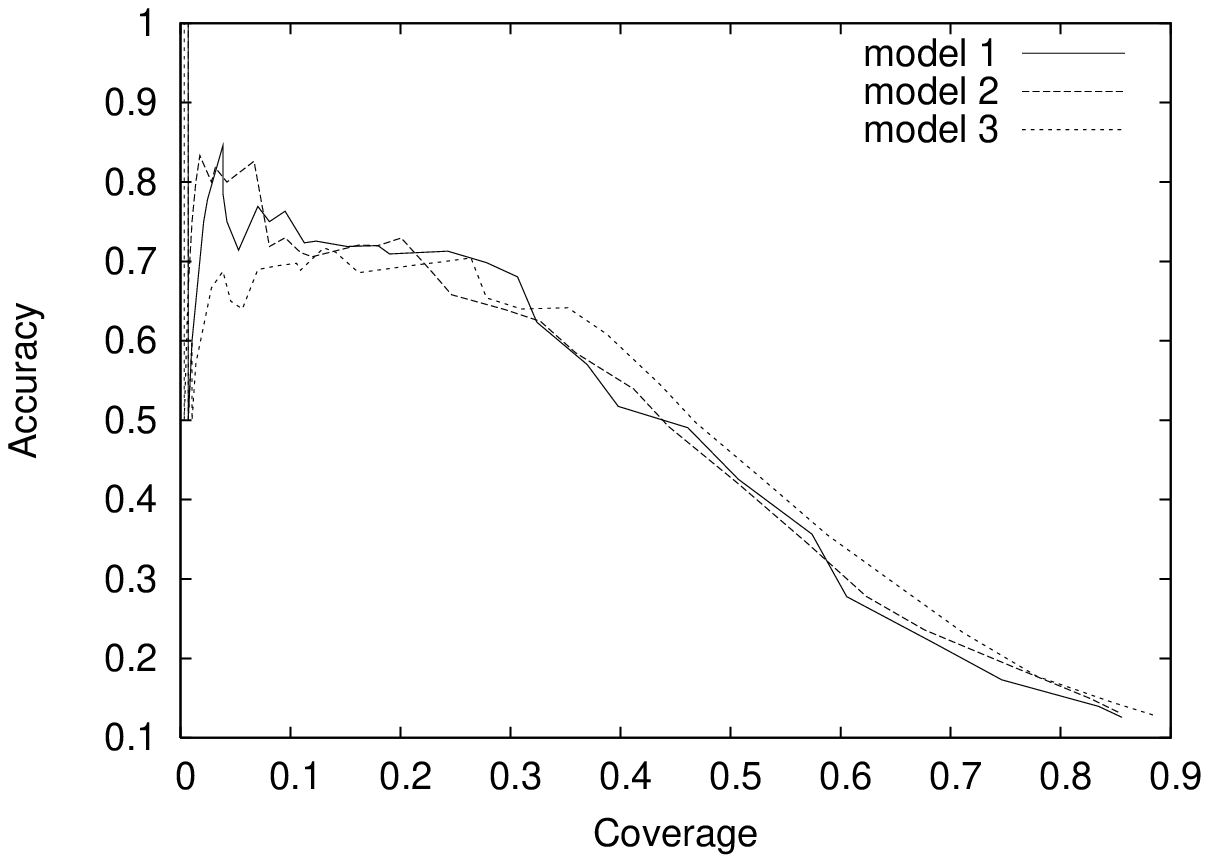}
\caption{Accuracy vs. coverage at a range of score thresholds for three
  homology models}
\label{ehnew-roc}
\end{center}
\end{figure}

Evaluation of promoter-prediction methods on a large scale is a difficult exercise,
since there are no large pieces of genomic sequence for which we can be certain
we know the complete set of transcribed regions, and even in the case on well-known
genes we often do not know the precise location at which transcription begins.
In \cite{down1}, we developed a pseudochromosome, derived from release 2.3 of
the chromosome 22 annotation.  As described above, this includes a subset of
284 experimentally verified gene structures.  The pseudochromosome was constructed
to include these genes while omitting all other annotated genes (which could
be substantially truncated).  We considered predictions (groups of one or more
overlapping windows which all have scores greater than some chosen threshold)
to be correct if they lie withing 2kb of an annotated gene start, and false otherwise.
Plotting accuracy (proportions of predictions which are correct) against
coverage (proportion of transcript starts which are detected by one of the
correct predictions) gives an ROC curve.  This is plotted for three different
models in figure \ref{ehnew-roc}.  Firstly, this shows that predictive
performance for all three models is rather similar similar.  It also shows
that they can function as accurate promoter predictors, with accuracy rising
to a plateau of around $0.7$.

\begin{figure}[!bth]
\begin{center}
\includegraphics[scale=0.66]{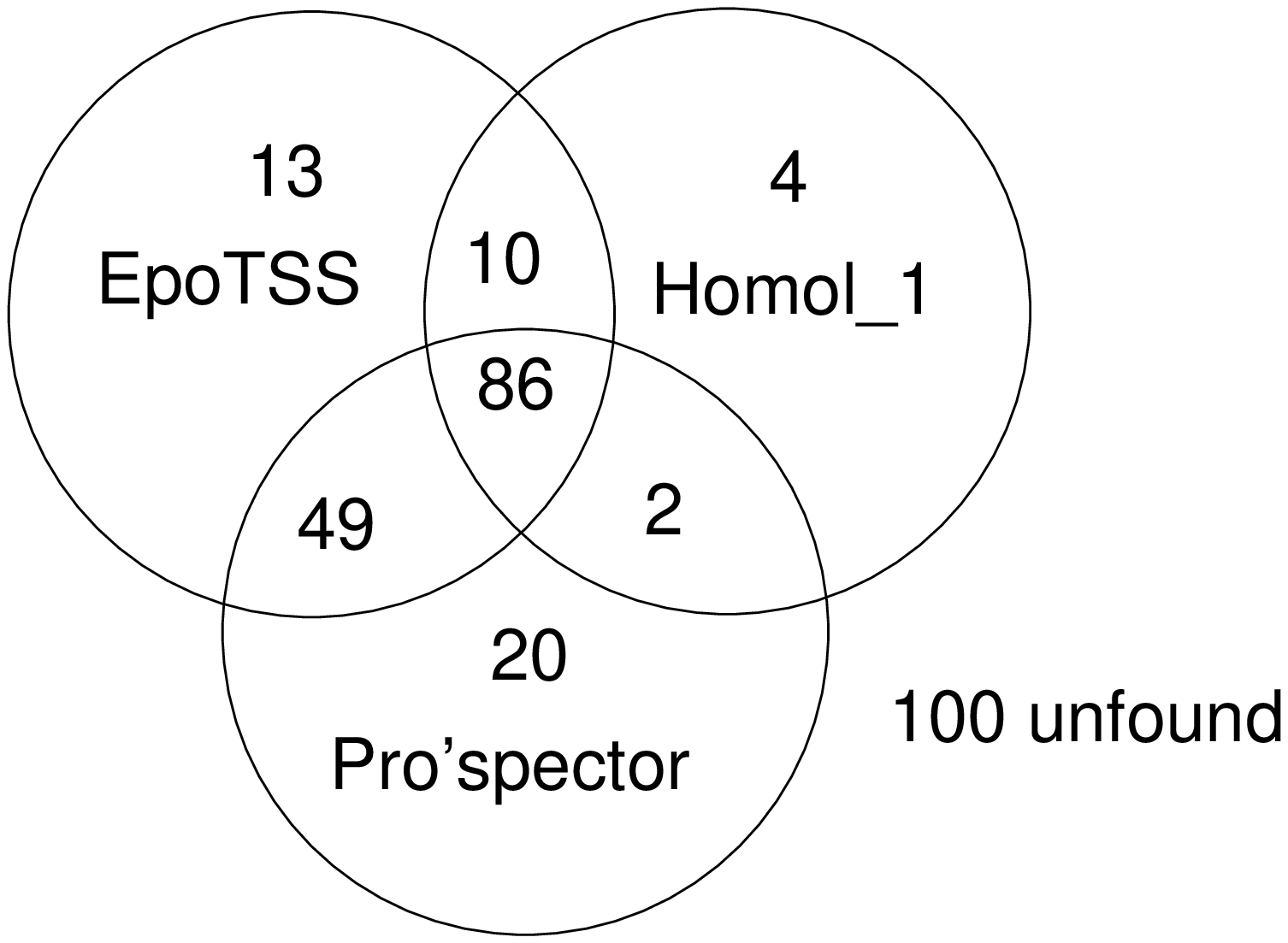}
\caption{Sets of pseudochromosome promoters correctly predicted by three
different prediction methods: EponineTSS \cite{down1} with a score threshold
of 0.999, PromoterInspector (labeled ``Pro'spector''), and the homology-EWS
model 1 with a score threshold of 0.91 (``Homol\_1'').}
\label{intersection}
\end{center}
\end{figure}

We picked model 1 for further study.  Using a score threshold of $0.91$, this gives
an accuracy of $0.68$ and a coverage of $0.31$.  We compared the set of genes
correctly detected by this model to two other methods: firstly, the EponineTSS
predictor described in \cite{down1}, and secondly, the published results from
the PromoterInspector program \cite{scherf.c22}.  PromoterInspector results were
mapped to pseudochromosome coordinates using the procedure described in
\cite{down1}.  Figure \ref{intersection} shows how the set of promoters detected
by these three distinct methods overlaps.  There are clearly strong correlations
between all three methods.  In particular, at this threshold the EpoHomol model
detects 98 promoters which were found by at least one of the other methods, but
only 4 novel promoters.

\section{Conclusions}

We have shown here that, when presented with a set of non-coding sequences which are
strongly conserved between human and mouse, a simple motif-oriented machine
learning system consistently builds models which are able to detect a
substantial fraction of human promoter regions with good accuracy.  This
strongly suggests that this promoter signal represents the most widely used
motif-based signal in functional non-coding sequence.  While the model learned
here can clearly be applied for the purpose of genome-wide promoter annotation, in practice existing
methods offer better coverage and (in the case of the EponineTSS predictor)
predictions for the precise location of the transcription start site.

It is interesting that the promoter model learned by this technique detected
substantially the same set of promoters as found by the EponineTSS and PromoterInspector
methods.  It has previously been remarked that these two methods detect similar
sets \cite{down1}, but this could perhaps be explained by the fact that both
methods were initially derived from similar sets of known promoter sequences (in both
cases, training data was extracted from the EPD database \cite{bucher.epd}.  In
the case of the homology models described here, there is no connection with EPD,
or any similar set of known promoters: the training data was picked purely
on the basis of its high similarity to corresponding portions of the mouse genome.
These result therefore support the alternate view that there is a particular
`easily detected' subclass of promoter sequences.

One distinct group of promoters, which previous results show may correspond to
this easily detected family, is those promoters associated with CpG islands [ref].
However, while a number of the motifs listed in table \ref{model.motifs} are G/C
rich and/or contain the CpG dinucleotide, by no means all of the motifs match
this description, and indeed one motif containing CpG has a negative weight in the
linear model -- their presence reduces the model output score -- while some A/T
rich motifs have positive weights.  We therefore believe that the signals detected
here are significantly more complex than a simple overrepresentation of CpG
dinucleotides.

\section{Materials and Methods}

\subsection{Genomic sequence and annotation}

Human genome sequence release NCBI33 and mouse genome release NCBIM30 were extracted
from Ensembl databases \cite{hubbard.ensembl}, which also contained gene predictions
from Genscan \cite{burge.genscan} and repeat data from RepeatMasker \cite{smit.repeatmasker}
and trf \cite{benson.trf}.  Curated annotation of gene
structures on human chromosome 6 was obtained from the Vega database \cite{vega}.
Vega and Ensembl data was extracted directly from the SQL databases using 
the BioJava toolkit with biojava-ensembl extensions \cite{biojava}.

\subsection{Genome alignments}

Human-mouse genome alignments were generate by the blastz alignment.  These were
subsequently re-scored and filtered to give a `tight' set of high-confidence
alignments, as described in \cite{schwartz.blastz}.  We downloaded the tight
alignment set from the UCSC genome website \cite{ucsc}.

\subsection{Pseudochromosome for testing promoter-finding methods}

A 16.3Mb pseudochromosome sequence was produced based on version 2.3 of the
curated annotation for human chromosome 22.  This includes all the experimentally-validated
gene structures and their upstream regions, while omitting regions containing
genes that are predicted but not fully verified.  In the case of a pair of
divergent genes where one has been verified and the second has not, their shared
upstream region was cut at the midpoint.  More information about
pseudochromosome construction is given in \cite{down1}.

\subsection{Eponine Windowed Sequence learning}

The Eponine Windowed Sequence (EWS) model is a machine learning system for identifying
a small set of motifs which can be effectively used to classify some set of training
sequences \cite{down.rvmseq}.  In this case, we applied a slightly restricted
version of the EWS trainer which omitted the ``Append Column'' sampling
rule, restricting the model to learning motifs with length less than or equal
to the length of the seed motifs.

\section{Acknowledgments}

Chromosome 22 annotation data version 2.3 were produced by the Chromosome 22 Annotation
Group at the Sanger Institute and were obtained from the World Wide Web at 
http://www.sanger.ac.uk/HGP/Chr22 (Dunham {\it et al.} unpublished data).  TD
would like to thank the Wellcome Trust for funding.

\end{document}